\documentclass[a4paper, 10pt, conference]{ieeeconf}      

\IEEEoverridecommandlockouts                              

\usepackage{amsfonts}
\usepackage{amsmath}
\usepackage{multirow}
\usepackage{booktabs} 
\usepackage{array}
\usepackage{graphicx} 
\usepackage{authblk} 

\begin{document}

\title{\LARGE \bf
LEAP: A Lightweight Encryption and Authentication Protocol \\ for In-Vehicle Communications
}


\author[1]{Zhaojun Lu}
\author[1]{Qian Wang}
\author[1]{Xi Chen}
\author[1]{Gang Qu}
\author[2]{Yongqiang Lyu}
\author[3]{Zhenglin Liu}
\affil[1]{University of Maryland, College Park, United States. \textit {\{lzj77521, qwang126, xchen128, gangqu\}@umd.edu}}
\affil[2]{Tsinghua University, Beijing, China. \textit {luyq@tsinghua.edu.cn}}
\affil[3]{Huazhong University of Science and Technology, Wuhan, China. \textit {liuzhenglin@hust.edu.cn}}

\maketitle
\thispagestyle{empty}
\pagestyle{empty}

\begin{abstract}

The Controller Area Network (CAN) is considered as the de-facto standard for the in-vehicle communications due to its real-time performance and high reliability. Unfortunately, the lack of security protection on the CAN bus gives attackers the opportunity to remotely compromise a vehicle. In this paper, we propose a Lightweight Encryption and Authentication Protocol (LEAP) with low cost and high efficiency to address the security issue of the CAN bus. LEAP exploits the security-enhanced stream cipher primitive to provide encryption and authentication for the CAN messages. Compared with the state-of-the-art Message Authentication Code (MAC) based approaches, LEAP requires less memory, is 8X faster, and thwarts the most recently proposed attacks.

\end{abstract}

\section{Introduction}

\subsection{Background}

Vehicle automation has been one of the fundamental applications in the field of Intelligent Transportation Systems (ITSs) \cite{R1}. With the demanding requirements of in-vehicle infotainment and driving safety, more and more advanced information processing technologies have been integrated into the modern vehicles. To support the functions of communication and vehicle control, a modern vehicle is a sophisticated and intelligent system with dozens of Electronic Control Units (ECUs) running software in the size of several hundred megabytes \cite{R2}. With numerous sensors, actors, and processors being connected on the in-vehicle networks, a vehicle can provide drivers and manufacturers with a variety of services, including vehicle diagnostics, Firmware Updating Over-the-Air (FOTA) \cite{R3}, and automatic driving, \textit{etc} \cite{R4}.

The Controller Area Network (CAN) protocol is the de-facto standard for the in-vehicle networks due to the dramatically decreased communication lines and the higher data transmission reliability \cite{R5}. Fig. 1 is the typical in-vehicle networks based on the CAN bus. The high-speed CAN bus has about $500\ Kbps$ data rate for the time-critical ECUs, while the low-speed CAN bus is used for other ECUs that have no real-time requirement. The original CAN protocol is designed under the assumption that all ECUs are legitimate, trustworthy, and operating according to their specifications \cite{R2}. No security consideration results in intrinsic vulnerabilities for the CAN protocol. Moreover, external interfaces such as the Second On-Board Diagnostic (OBD-II), Bluetooth, Wi-Fi, and the Global Positioning System (GPS) provide opportunities for attackers to break into the unprotected CAN bus. Koscher \textit{et al.} \cite{R6} demonstrated that it was possible to control the entire automotive electronics system via the available interfaces. In 2015, this threat became a reality when Miller and Valasek \cite{R7} demonstrated how to remotely control a Jeep Cherokee by compromising the vulnerable ECUs, which triggered Chrysler to recall approximately 1.4 million vehicle. 

\begin{figure}[t]
  \centering
  \includegraphics[width=\linewidth]{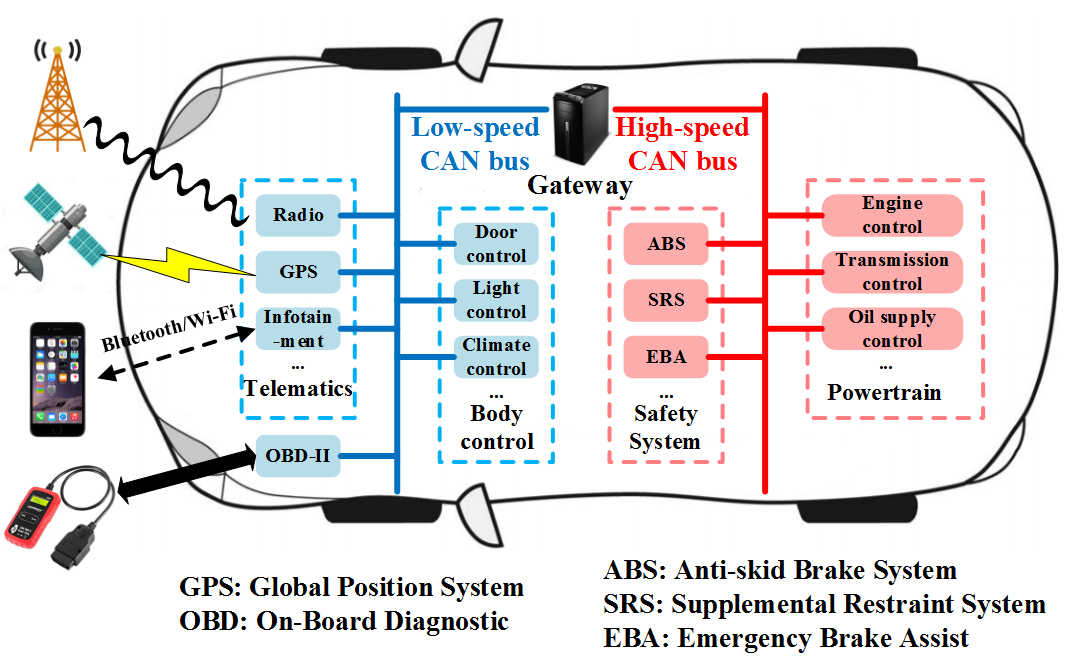}
  \caption{In-vehicle networks based on the CAN bus.}
\end{figure}

\subsection{Current Approaches}

The current approaches to secure the CAN bus can be classified into two categories, \textit{i.e.}, the Intrusion Detection Systems (IDSs) and the authenticated CAN \cite{R8}. The anomaly-based IDSs focus on the features of the CAN bus to detect the known and unknown attacks \cite{R9}. The most essential issue to be considered when designing an anomaly-based IDS is what feature the detector exploits to discriminate between normal and abnormal. Since there is no timestamp carried in the CAN messages, Cho \textit{et al.} \cite{R10} developed the Clock-based IDS (CIDS) that fingerprints the sender ECUs according to the estimated clock skews. Considering the nature that most of the CAN messages on the CAN bus are cyclic and the format of the data-field is fixed due to rigid signal allocation, the low-dimensional features of the CAN messages could be explored to detect the anomaly on the CAN bus \cite{R11}. The downside is that no single IDS is able to detect all the attacks. Without protecting the confidentiality and authentication of the CAN messages, the sophisticated attackers can bypass a specific type of IDS and compromise the in-vehicle networks \cite{R12}.

The authenticated CAN suggests to attach a Message Authentication Code (MAC) to each CAN message for authentication. MAC protects a message's authenticity and allows the verifiers who also possess the secret key to detect any changes in the message content. Woo \textit{et al.} \cite{R13} designed an MAC-based security protocol that inserted the divided MAC in the extended identifier field and the Cyclic Redundancy Check (CRC) field. However, the intensive computation of MAC could degrade or even cripple an ECU's normal operation. The Denial of Service (DoS) attacks can be easily launched on ECUs where the MAC-based security is implemented \cite{R14}.

\subsection{Contributions}

There are three major challenges to design a practical security scheme for CAN. First, the confidentiality and authentication are essential for each CAN message to resist the known and unknown attacks. Second, the resource of most life-critical ECUs is limited to support the simple function. The security scheme should guarantee the real-time performance of the CAN bus. Third, because a vehicle has a life cycle of more than five years, the compatibility of the in-vehicle networks should be considered. Physical modification of the in-vehicle architecture or replacement of a large number of ECUs are not acceptable.

In this paper, we present the basic attack models based on the thorough analysis of the recently proposed attack cases. Then we propose a Lightweight Encryption and Authentication Protocol (LEAP) to provide the in-vehicle networks with security, efficiency, and compatibility. The major contributions of our work are as follows:

\begin{itemize}
\item LEAP exploits the stream cipher RC4 instead of MAC to encrypt and authenticate each CAN message, which is lightweight in terms of time consumption and storage overhead. Thus, the efficiency and real-time performance of the CAN bus can be guaranteed. Moreover, LEAP only modifies the software of ECUs so that it is compatible with the existing in-vehicle networks. 
\item Considering that RC4 is not as secure as the classical cryptographic algorithms, we design a key management mechanism with low cost and high efficiency to enhance the security of LEAP. The session keys will be updated and distributed to ECUs periodically to resist the brute-force attack. This process will be performed when the vehicle is at idle speed to eliminate the impact on the CAN bus communications.
\item A set of experiments has been conducted on the same platform as \cite{R10}. We launch the basic attack models to demonstrate the security of LEAP. The results show that LEAP consumes less memory and is of 8X higher efficiency than the MAC-based approaches. 
\end{itemize}

The remainder of this paper is organized as follows. Section II gives a brief description of CAN and analyzes the attack models systematically. The proposed LEAP is elaborated in Section III. Section IV discusses how the proposed approach can thwart different attacks. Performance evaluation is reported in Section V before we conclude in Section VI.

\section{Attack Models}

\begin{figure}[t]
  \centering
  \includegraphics[width=\linewidth]{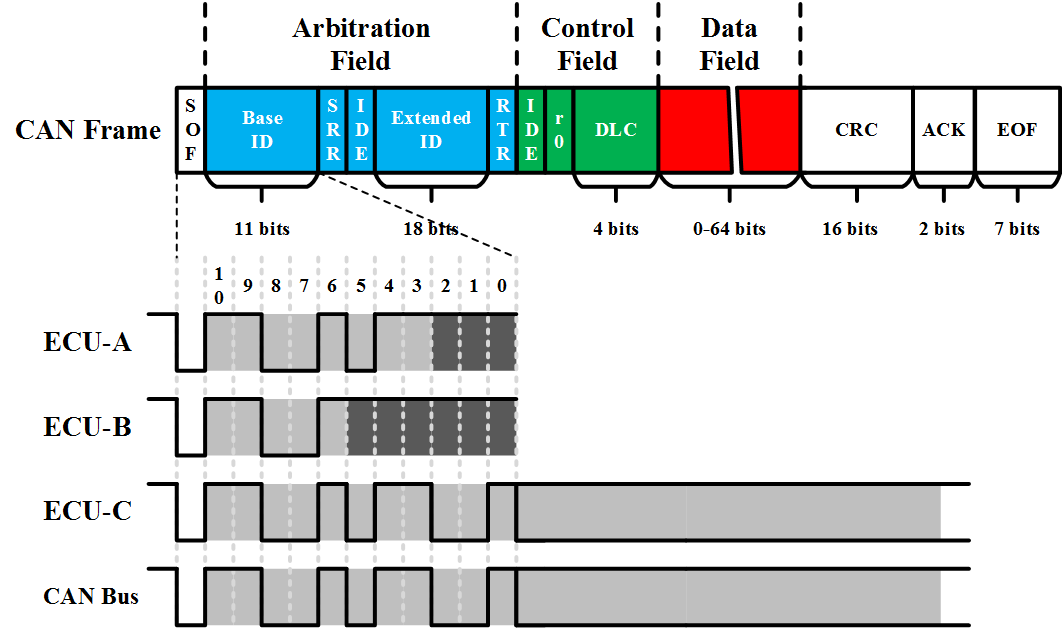}
  \caption{Format of the CAN message and the bus arbitration.}
\end{figure}

\begin{figure*}[t]
  \centering
  \includegraphics[width=0.85\linewidth]{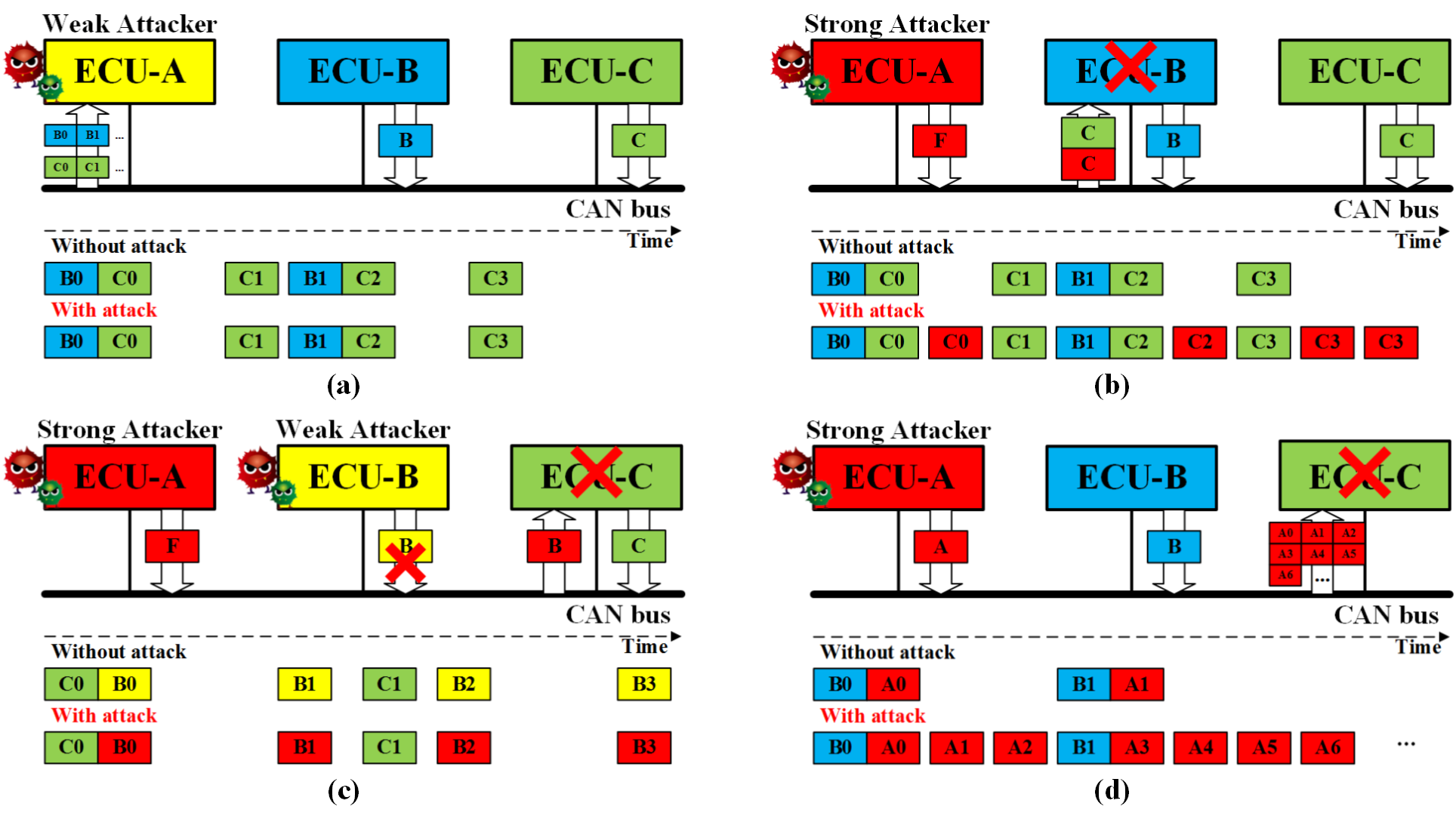}
  \caption{The basic attack models. (a) Eavesdrop Attack. (b) Replay Attack. (c) Masquerade Attack. (d) Flooding Attack.}
\end{figure*}

CAN messages are sent by ECUs in compliance with the format as shown in Fig. 2. In addition to the Data Field, each message carries necessary information for the bus arbitration and the error management. A CAN message contains a unique Identifier (ID) representing its source and priority without other information about the sender or the receiver. Fig. 2 illustrates a situation that three ECUs compete to get access to the shared bus. The Data Length Code (DLC) contains the length of the Data Field which is in the range of 0 to 64 $bits$. This message format and the underlying CAN protocol have four major security vulnerabilities \cite{R6}: the broadcast nature makes the messages available to all ECUs on the CAN bus; it will be vulnerable to DoS attacks due to the simple arbitration mechanism; no authentication field to identify the sender and the receiver of each message; and the access control is non-existing or very weak.

According to the attackers' capabilities, they can be considered as either {\bf strong attackers} or {\bf weak attackers} \cite{R15}. Strong attackers are those who can fully compromise an ECU and gain control of its software and memory. It means that a strong attacker is able to access any critical data stored in the memory. Thus, a strong attacker not only can inject messages with arbitrary \textit{IDs} and contents on the CAN bus, but also can disable the security mechanisms built in ECUs. Weak attackers are those who cannot access critical data in the compromised ECUs. Therefore, a weak attacker can only stop the weakly compromised ECU or eavesdrop on the CAN bus. In this paper, we focus on the existing attack cases that have been implemented to severely impair the in-vehicle security \cite{R7,R13,R15}. These attacks are composed of the following four basic attack models:

\begin{itemize}
\item {\bf Eavesdrop Attack.} As shown in Fig. 3(a), the first step of the attacks is to eavesdrop on the CAN bus to figure out the exact format and content to control the target ECU. The lack of confidentiality for the CAN messages makes it very easy for a weak attacker ECU-A to launch the eavesdrop attack.
\item {\bf Replay Attack.} As shown in Fig. 3(b), a strong attacker ECU-A sends a message received from the CAN bus without modifying it. Due to the lack of authentication on the CAN bus, the receiver ECU-B will function abnormally under the replayed control information.
\item {\bf Masquerade Attack.} Having known the content and frequency of the messages from ECU-B, the strong attack could launch the masquerade attack in Fig. 3(c). The attacker first suspends the message transmission of ECU-B, then uses the fully compromised ECU-A sends messages using ECU-B's \textit{ID} to manipulate other legitimate ECUs.
\item {\bf Flooding Attack.} As shown in Fig. 3(d), the flooding attack is a kind of DoS attack, whose objective is to exhaust the target ECU's computational resources by sending it large amounts of messages.
\end{itemize}

\section{Lightweight Encryption and Authentication Protocol}

LEAP consists of two mechanisms, \textit{i.e.}, the key management mechanism, and the message encryption and authentication mechanism. In the key management mechanism, a Secure ECU is responsible to generate the session keys periodically and distribute them to each pair of communicating ECUs. In the message encryption and authentication mechanism, each pair of communicating ECUs use the shared session key to guarantee the confidentiality and authentication of the CAN messages. The notations and definitions are listed in Table I.

\begin{table}[t]
	\newcommand{\tabincell}[2]{\begin{tabular}{@{}#1@{}}#2\end{tabular}}
	\footnotesize
	\caption{Notations and Definitions}
	\begin{tabular*}{0.9\linewidth}{@{\extracolsep{\fill}}ll}
		\toprule			
		\bf{Notations}      & \bf{Definitions}\\
		\midrule			
		\midrule			
		\textit{ID}$_i$     & \textit{ID} of ECU$_i$ \\
		\midrule			
		\textit{CTR}$_{ij}$ & Massage counter between ECU$_i$ and ECU$_j$ \\
		\midrule			
		$\mathcal{LK}_i$    & Long-term symmetric key of ECU$_i$ \\
		\midrule			
		$\mathcal{SK}^n_{ij}$  & $n$-th session key of ECU$_i$ and ECU$_j$ \\
		\midrule			
		\textit{PM}         & Plaintext of the CAN message \\
		\midrule			
		\textit{CM}         & Ciphertext of the CAN message \\
		\midrule			
		$\mathcal{K}$       & Key stream generated by RC4 using $\mathcal{SK}$\\
		\midrule			
		$\texttt{KDF}_{key}$() & Keyed one-way function for key derivation\\
		\midrule			
		$\texttt{H}_{key}$() & Keyed hash function to generate MAC \\
		\midrule
		$\texttt{E}_{key}$()  & Symmetric encryption algorithm \\
		\midrule			
		$\texttt{D}_{key}$()  & Symmetric decryption algorithm \\
		\toprule			
	\end{tabular*}
\end{table}

\subsection{Key Management Mechanism}

\begin{figure}[t]
  \centering
  \includegraphics[width=0.8\linewidth]{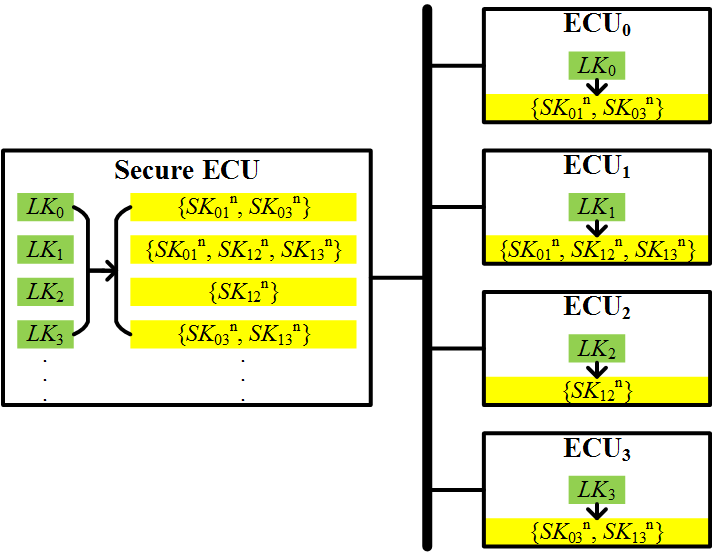}
  \caption{Key management mechanism.}
\end{figure}

The key management mechanism is of top priority to meet all the security requirements. As shown in Fig. 4, the long-term symmetric keys $\mathcal{LK}_i$ are generated in the vehicle manufacturing stage or when an ECU is replaced. The session keys $\mathcal{SK}^n_{ij}$ are derived from $\mathcal{LK}_i$ and $\mathcal{LK}_j$ periodically for the message encryption and authentication mechanism. The sets of session keys are different because each ECU communicates with specific ECUs. For example, ECU$_0$ only communicates with ECU$_1$ and ECU$_3$. The security of the session keys is enhanced with two methods. First, $\mathcal{SK}^n_{ij}$ are only shared between ECU$_i$ and ECU$_j$. Thus, a fully compromised ECU cannot deceive ECUs that do not directly communicate with it. Second, $\mathcal{SK}^n_{ij}$ are updated periodically to resist the brute-force attack.

All the long-term symmetric keys are stored in the Secure ECU, which has sufficient resource and high security level and acts as the centralized manager to secure the key management mechanism. We optimize the key update and distribution process proposed in \cite{R13} to achieve same security with higher efficiency. The Advanced Encryption Standard (AES) is used as the symmetric cryptographic algorithms ($\texttt{E}_{key}$ and $\texttt{D}_{key}$). The Secure Hash Algorithm (SHA) is used as the keyed hash function ($\texttt{H}_{key}$). ECU$_i$ and ECU$_j$ are a pair of communicating ECUs, Fig. 5 illustrates the six steps to update and distribute the shared session key to them.

{\bf Step 1.} The Secure ECU selects a random value as $Seed^n_{ij}$, then generates the symmetric session key $\mathcal{SK}^n_{ij}$ using the long-term symmetric key $\mathcal{LK}_{i}$ and $\mathcal{LK}_{j}$.

{\bf Step 2.} The Secure ECU encrypts $\mathcal{SK}^n_{ij}$ using $\mathcal{LK}_{i}$ and $\mathcal{LK}_{j}$ respectively to generate $Cipher_i$ and $Cipher_j$.

{\bf Step 3.} The Secure ECU generates \textit{MAC-1}$_i$ of the concatenation of \textit{ID}$_i$ and $Cipher_i$ using $\mathcal{LK}_{i}$. A key-update request containing \textit{ID}$_i$, \textit{MAC-1}$_i$ and $Cipher_i$ is broadcasted for ECU$_i$. The same process applies to ECU$_j$.

{\bf Step 4.} ECU$_i$ verifies \textit{MAC-1}$_i$ to authenticate the key-update request. Then ECU$_i$ gets $\mathcal{SK}^n_{ij}$ by decrypting $Cipher_i$ using $\mathcal{LK}_{i}$. The same process applies to ECU$_j$.

{\bf Step 5.} ECU$_i$ generates \textit{MAC-2}$_i$ of \textit{ID}$_i$ using $\mathcal{SK}^n_{ij}$ and sends the key-update response to the Secure ECU for authentication. The same process applies to ECU$_j$.

{\bf Step 6.} The Secure ECU verifies \textit{MAC-2}$_i$ and \textit{MAC-2}$_j$ to authenticate the key-update response from ECU$_i$ and ECU$_j$.

We assume the security of AES and SHA, as well as the long-term symmetric keys. \cite{R13} gives the strict security proof of the key update and distribution process. Therefore, $\mathcal{SK}^n_{ij}$ can be securely transmitted from the Secure ECU to ECU$_i$ and ECU$_j$.

\begin{figure}[t]
  \centering
  \includegraphics[width=\linewidth]{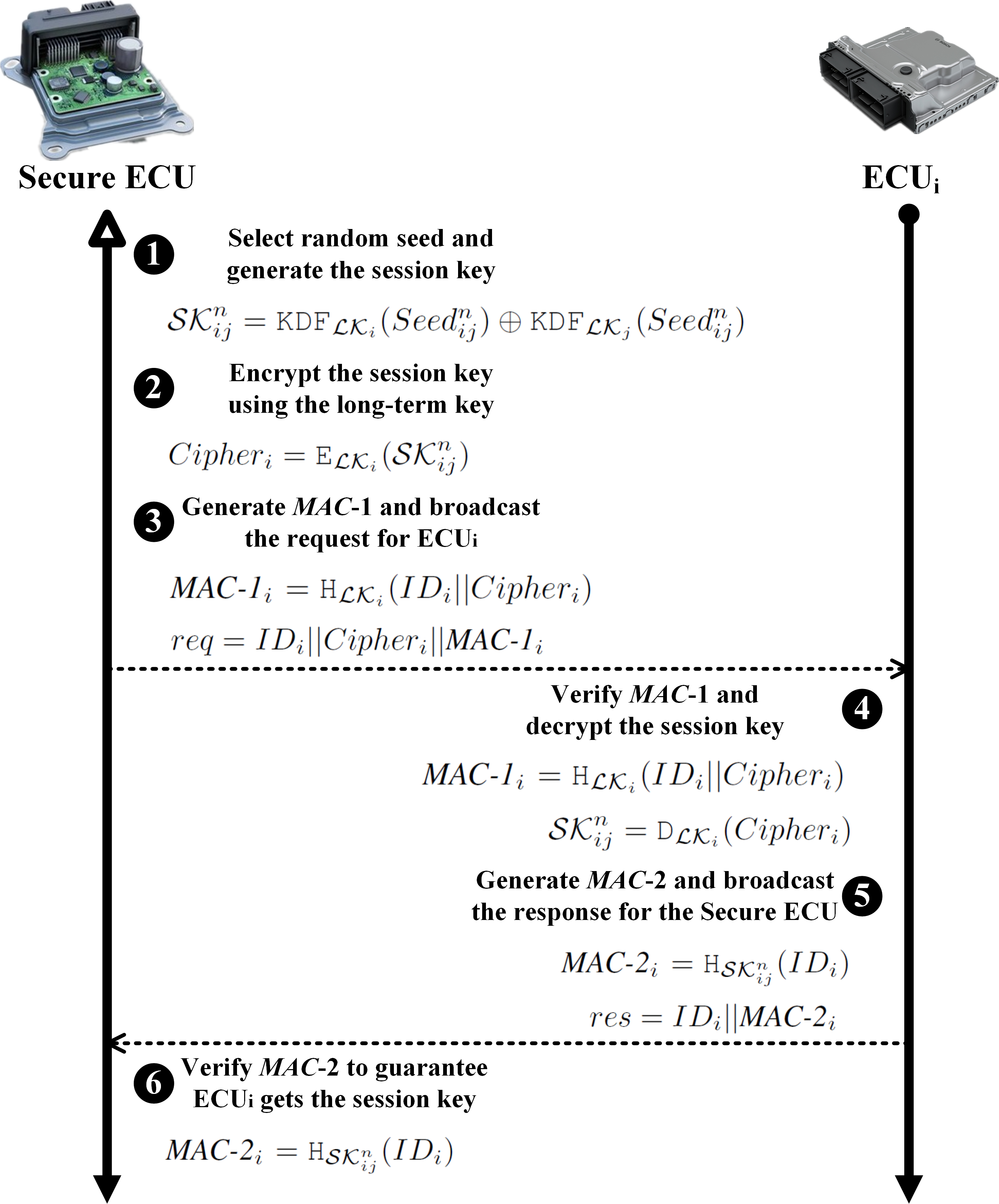}
  \caption{Session key update and distribution process.}
\end{figure}

\subsection{Message Encryption and Authentication Mechanism}

RC4 is one of the most widely-used stream cipher algorithms that consists of the Key Scheduling Algorithm (KSA) and the Pseudo-Random Generation Algorithm (PRGA) \cite{R16}. KSA generates the initial permutation from a key with $m\ bytes$. Typically, $m$ is in the range of 5 to 64. The main part of RC4 is PRGA that produces one-byte output in each step. The encryption is an XOR of the pseudo-random sequence with the plaintext.

After the key update and distribution process, each pair of communicating ECUs will have a symmetric session key to perform the message encryption and authentication process based on RC4 instead of the AES and MAC-based approaches. The sender ECU$_i$ and the receiver ECU$_j$ are a pair of communicating ECUs, the proposed message encryption and authentication approach is illustrated in Fig. 6. There are four steps for ECU$_i$ before broadcasting a CAN message:

\begin{figure}[b]
  \centering
  \includegraphics[width=0.8\linewidth]{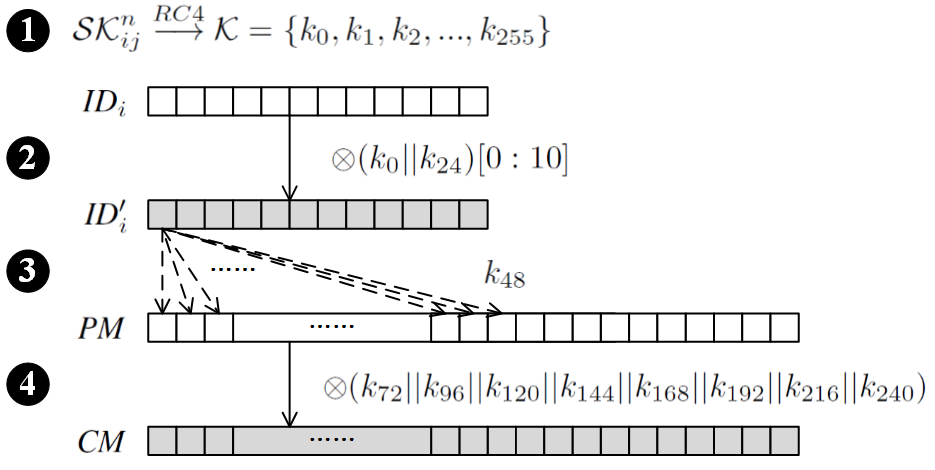}
  \caption{RC4 based encryption and authentication for the CAN message.}
\end{figure}

{\bf Step 1.} ECU$_i$ performs RC4 to generate the key stream $\mathcal{K} = \{k_0, k_1, k_2,..., k_{255}\}$ using $\mathcal{SK}^n_{ij}$.

{\bf Step 2.} ECU$_i$ performs an XOR operation between ECU$_i$'s \textit{ID}$_i$ and $(k_0||k_{24})[0:10]$ to generate \textit{ID}$'_i$.

{\bf Step 3.} ECU$_i$ determines the location to insert \textit{ID}$'_i$ in the Data Field according to $k_{48}$. Since \textit{ID}$'_i$ has 11 $bits$ and the Data Field has 64 $bits$, there are 54 possible locations for \textit{ID}$'_i$. We set \textit{PM}$[l] (l = k_{24}\ mod\ 54)$ as the location of \textit{ID}$'_i[0]$.

{\bf Step 4.} ECU$_i$ performs an XOR operation between \textit{PM} and $(k_{72}||k_{96}...||k_{240})$ to generate \textit{CM}. Finally, ECU$_i$ broadcasts the CAN message with \textit{CM} in the Data Field.

Since the proposed approach will generate distinct key stream $\mathcal{K}$ for each CAN message, the sender ECU$_i$ and the receiver ECU$_j$ jointly manage a message counter \textit{CTR}$_{ij}$ for synchronization. The 5-bit \textit{CTR}$_{ij}$ is stored in the pair of communicating ECUs and will be reset in the key update and distribution process. \textit{CTR}$_{ij}$ will add 1 if the sender ECU$_i$ broadcasts a CAN message or the receiver ECU$_j$ successfully receives a CAN message. When receiving a CAN message, the receiver ECU$_j$ first generates the key stream $\mathcal{K}$ using the shared symmetric session key $\mathcal{SK}^n_{ij}$ according to \textit{CTR}$_{ij}$. Then the receiver ECU$_j$ decrypts \textit{CM} using the key stream $\mathcal{K}$ to get \textit{PM} and \textit{ID}$_i$. The receiver compares \textit{ID}$_i$ out of \textit{CM} with \textit{ID} of the received message. If they are same, it is verified that this CAN message is from ECU$_i$. If the authentication fails, this message will be discarded by ECU$_j$ and \textit{CTR}$_{ij}$ will remain unchanged.

\begin{figure}[b]
  \centering
  \includegraphics[width=0.8\linewidth]{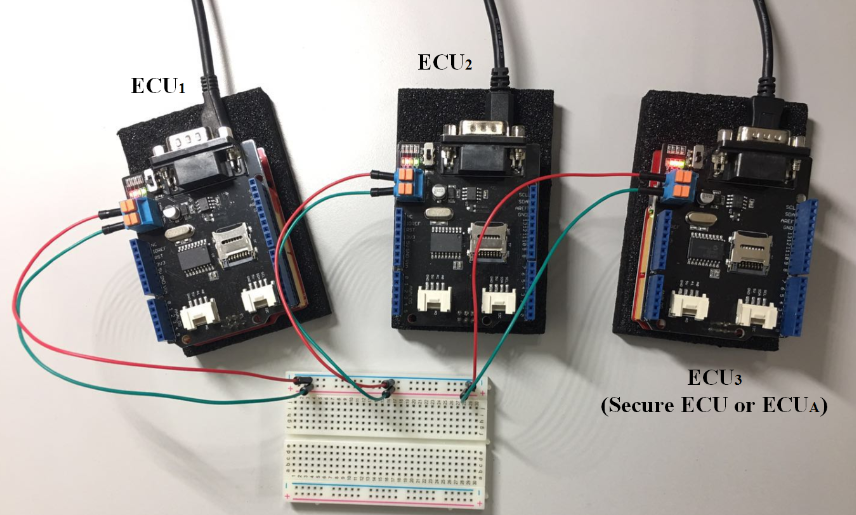}
  \caption{CAN bus prototype.}
\end{figure}

LEAP acts as the last line of defense to protect the confidentiality and authentication of the CAN messages. It prevents a compromised ECU from eavesdropping on the CAN bus or using other ECUs' \textit{IDs} to broadcast messages. The only imperfection of our design is that a fully compromised ECU$_i$ can still deceive ECU$_j$ by sending fake messages using the shared symmetric session key $\mathcal{SK}^n_{ij}$.

\section{Security Analysis}

In this section, we first provide the experiment settings. Then, we measure the cost of the brute-force attack on RC4 to explain how to guarantee the security of the session keys. Finally, we launch the basic attack models and analyze the security of LEAP theoretically and experimentally.

\subsection{Experiment Settings}

As shown in Fig. 7, we configure a CAN bus prototype similar to \cite{R10}, in which there are three ECUs connecting to each other via a 2-wire bus. ECU$_1$ (\textit{ID}$_1$ $0x0001$) and ECU$_2$ (\textit{ID}$_2$ $0x0002$) is a pair of communicating ECUs. ECU$_3$ acts as the Secure ECU (\textit{ID}$_S$ $0x0000$) or a fully compromised ECU$_A$ (\textit{ID}$_A$ $0x0003$). Each ECU consists of a Seeeduino v4.2 board and a SeeedStudio CAN bus shield v2.0. Seeeduino board is equipped with an ATmega328 microcontroller containing $16\ MHz$ clock, $32\ KB$ flash, $2\ KB$ RAM, and $1\ KB$ EEPRAM. CAN bus shield adopts MCP2515 CAN bus controller with SPI interface and MCP2551 CAN transceiver to give Seeeduino CAN bus communication capability.

\subsection{Security of RC4}

\begin{table}[t]
    \newcommand{\tabincell}[2]{\begin{tabular}{@{}#1@{}}#2\end{tabular}}
    \footnotesize
    \centering
    \caption{Cost of brute-force attack on RC4}
    \label{Table2}
    \begin{tabular}{ccccc}
    \toprule
    \multirow{2}{*}{ } &  \multirow{2}{*}{\tabincell{c}{Time for each \\ key search / $\mu s$}} & \multicolumn{3}{c}{\tabincell{c}{Average attack time / $hour$ \\ for different key lengths}}  \\
    \cmidrule(r){3-5} 
    & & $40\ bits$ & $64\ bits$ & $128\ bits$\\
    \midrule
    Laptop &  $30$  &  $4,581$ & $7.7 \times 10^{10}$ & $1.4 \times 10^{30}$ \\
    \specialrule{0em}{3pt}{3pt}
    FPGA   &  $0.1$ &  $15$ & $2.5 \times 10^{8}$ & $4.7 \times 10^{27}$ \\
    \bottomrule
    \end{tabular}
\end{table}

The security of LEAP is based on two prerequisites. First, the key update and distribution process is secure. Second, the session key $\mathcal{SK}^n_{ij}$ for RC4 cannot be cracked in one update cycle. Since $\mathcal{SK}^n_{ij}$ is generated by the Secure ECU and is encrypted using AES, the confidentiality of $\mathcal{SK}^n_{ij}$ are guaranteed. The 32-bit truncated MAC is used to authenticate $\mathcal{SK}^n_{ij}$ with the secure long-term symmetric keys. Thus, $\mathcal{SK}^n_{ij}$ is secure from the Secure ECU to ECU$_i$ and ECU$_j$ \cite{R13}. 

Although RC4 has been broken in the transport layer security protocol \cite{R17}, it is still sufficiently secure to be deployed in the CAN bus as long as its key-space is large enough against the brute-force attack. As explained in Section II, a compromised ECU$_A$ is able to passively eavesdrop on the CAN bus. Considering the worst case, the compromised ECU$_A$ knows the plaintext from the target ECU. Thus, it is a kind of known-plaintext attack. The attacker chooses a candidate key $\mathcal{SK}$ from the key-space $\mathbb{K}$ to operate the decryption then verifies the correctness of $\mathcal{SK}$ by comparing with the plaintext. If the key of RC4 is $n$-bit, the scale of $\mathbb{K}$ is $2^n$ and it will take on an average of $2^{n-1}$ iterations to get the correct key. Therefore, the time complexity of the brute-force attack on RC4 is $O(2^n)$. We run the brute-force attack on both a laptop with $2.5\ GHz$ CPU and $8.00\ GB$ RAM and a FPGA with $100\ MHz$ clock. The cost of the brute-force attack on RC4 with different key length is listed in Table II. For example, for a 40-bit key, it takes $4,581\ hours$ for laptop or $15\ hours$ for FPGA to crack RC4. Even though GPU can be used to accelerate the brute-force attack \cite{R18}, the results show that the 128-bit session key is secure enough for RC4.

\subsection{Security of LEAP}

We first theoretically prove the security of LEAP. Then, ECU$_A$ act as the attacker to experimentally test the resistance of LEAP against each basic attack model. For simplicity, the 128-bit session key is set as $\{0x00, 0x01,..., 0x0f\}$, the content of a CAN message is set as $\{0x0a, 0x0b, 0x0c, 0x0d, 0x0e, 0x0f\}$ (two bytes are left to insert \textit{ID}$'$). The data flow of the four basic attack models is shown in Fig. 8. The security of LEAP is guaranteed by the following lemmas:

\begin{figure}[t]
  \centering
  \includegraphics[width=0.8\linewidth]{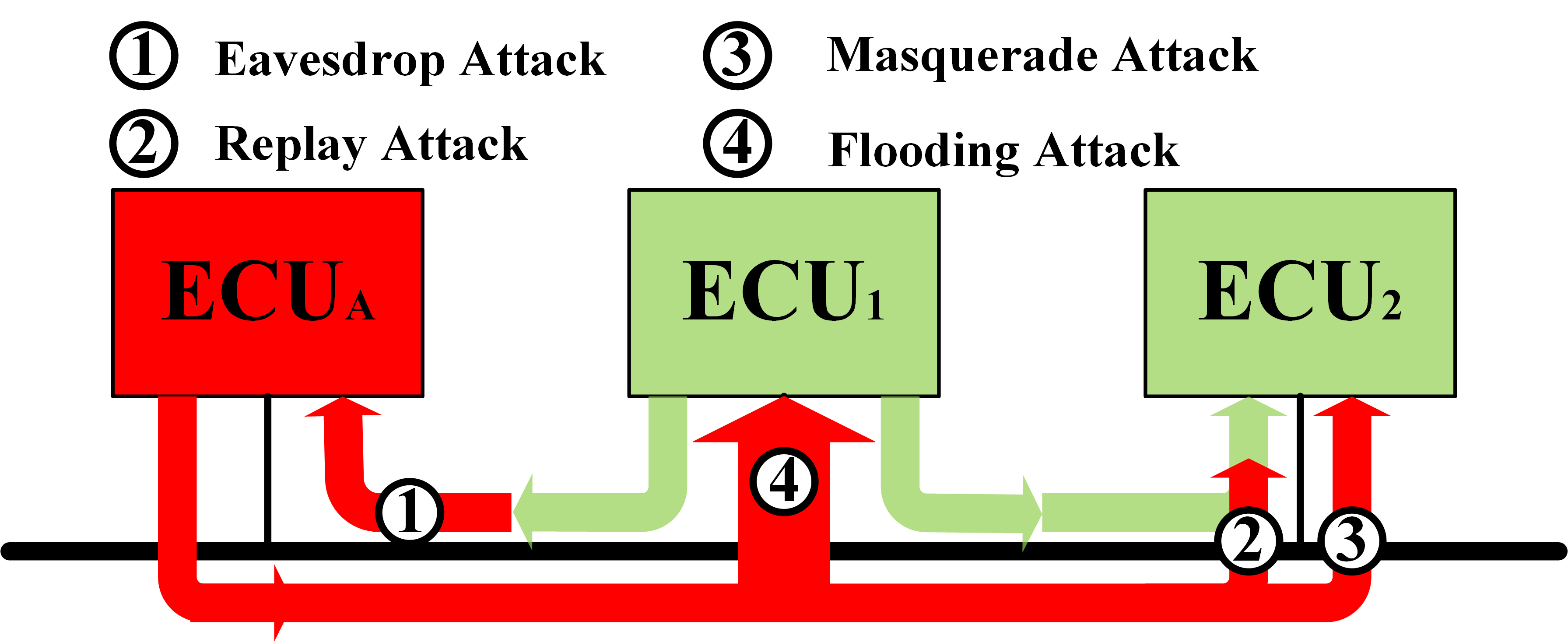}
  \caption{Data flow of basic attack models.}
\end{figure}

{\bf Lemma 1.} LEAP is secure against the eavesdrop attack.

{\bf Proof.} In LEAP, the 64-bit payload is encrypted using the key-stream generated by RC4. Vanhoef \textit{et al.} \cite{R17} present an efficient plaintext recovery attack on RC4 that can decrypt a secure cookie with a success rate of $94\%$ using $9\times2^{27}$ ciphertexts in $75\ hours$. Since the most frequency of the CAN messages is about $200\ Hz$, an ECU can send $200\times3,600\times24 = 17,280,000$ messages in $24\ hours$, which is much less than $9\times2^{27}$. Thus, LEAP is resistant against the eavesdrop attack.

In the experiment, ECU$_1$ broadcasts 10 CAN messages per second and ECU$_A$ gets all messages from ECU$_1$. In this case, the encrypted CAN messages with same content are $\{0x556ecf97d32e7c6a, 0xb1df00bff164ecd1, 0xb2cd8430\\0be83bea, 0xac8f35751325a6dd, 0x9823d8520d0f9235...\}$. It is difficult for the attacker to recover the plaintext and the session key from the ciphertexts in $24\ hours$.

{\bf Lemma 2.} LEAP is secure against the replay attack.

{\bf Proof.} A counter is kept for each pair of communicating ECUs for synchronization. The sender will generate distinct key-stream for each message according to the counter. Therefore, the replay attack cannot be successfully launched as long as the session key is secure.

ECU$_1$ broadcasts 10 CAN messages per second, ECU$_A$ replays all messages received from ECU$_1$ for $10,000\ times$, and ECU$_2$ receives messages from ECU$_1$ and ECU$_A$. For example, ECU$_2$ decrypts the first 10 replayed messages from ECU$_A$ to get the \textit{IDs} $= \{0x1cf3, 0x23c1, 0x9e1e, 0xee5e, 0x8c36, 0x521b, 0x6682, \\0xa194, 0x3cbe, 0x58a3\}$. Since all the \textit{IDs} from the $10,000$ replayed messaged are different from \textit{ID}$_1$ $0x0001$, these replayed messages will be discarded by ECU$_2$.

{\bf Lemma 3.} LEAP is secure against the masquerade attack.

{\bf Proof.} LEAP provides authentication for each message by encrypting and inserting the sender's \textit{ID} into the Data Field. Without knowing the session key, the masquerade attack will be thwarted because the \textit{ID} out of the Data Field will be different with the \textit{ID} of the messages from the attacker.

ECU$_A$ uses ECU$_1$'s \textit{ID}$_1$ $0x0001$ to broadcast messages without knowing the session keys and ECU$_2$ receives messages from ECU$_A$. The result is same as the replay attack that \textit{ID} decrypted out of the Data Field does not equal to ECU$_1$'s \textit{ID}$_1$. Thus, ECU$_2$ will discard the messages from ECU$_A$.

{\bf Lemma 4.} LEAP is secure against the flooding attack.

{\bf Proof.} In the flooding attack, the attacker attempts to exhaust the target ECU's computational resource by sending it huge amounts of messages. If the attacker uses its real \textit{ID}, the receiver can easily detect the abnormal frequency and alert that the CAN bus is under the flooding attack. If the attacker uses other ECUs' \textit{IDs}, the authentication will fail and the counter of the receiver will not increase, which means the receiver will not generate new key-stream for the subsequent messages. Since the decryption with the current key-stream merely involves several XOR operations, even if the attacker occupies full bandwidth of CAN, the number of messages that the attacker sends on the CAN bus is less than that the receiver can decrypt per second.

Considering the worst case, ECU$_A$ occupies full bandwidth of CAN ($500\ Kbps$) to launch the flooding attack to ECU$_2$. ECU$_2$ has to decrypt all the received messages. The results show that ECU$_2$ is able to decrypt and authenticate more than $12,000$ CAN messages per second while ECU$_A$ can send at most $4,400$ CAN messages per second on the CAN bus. Thus, ECU$_2$ can operate normally under the flooding attack

\section{Performance Evaluation}

LEAP consists of two processes, the key update and distribution process and the encryption and authentication process for the CAN messages. We set the CAN bus prototype at $500\ Kbps$ as the high-speed in-vehicle CAN bus. We first measure the cost of the Secure ECU (ECU$_3$) and the general ECUs (ECU$_1$ and ECU$_2$) in the key update and distribution process. Then we compare LEAP with the state-of-the-art MAC-based approaches in the message encryption and authentication process.

\subsection{Cost of Key Update and Distribution Process}

\begin{table}[b]
    \newcommand{\tabincell}[2]{\begin{tabular}{@{}#1@{}}#2\end{tabular}}
    \footnotesize
    \centering
    \caption{Cost of key update and distribution process}
    \label{Table2}
    \begin{tabular}{ccccc}
    \toprule
    \multirow{2}{*}{ }  & \multicolumn{2}{c}{Memory / $byte$} &  \multirow{2}{*}{\tabincell{c}{Processing \\ time / $ms$}} & \multirow{2}{*}{\tabincell{c}{Total \\ time / $ms$}}\\
    \cmidrule(r){2-3} 
    & Code memory & RAM\\
    \midrule
    Secure \\ ECU & $7,880$ & $640$ & $70.2$ & \multirow{3}{*}{$199.4$} \\
    \specialrule{0em}{3pt}{3pt}
    General \\ ECU & $5,660$ & $419$ & $15.9$  \\
    \bottomrule
    \end{tabular}
\end{table}

As shown in Fig. 5, a general ECU performs one decryption of AES-128 to get $\mathcal{SK}^n_{ij}$ and two keyed hash functions for the authentication. The payload of the key-update request is $128 + 32 = 160\ bits$ so that three CAN messages are required. The interval between two consecutive CAN messages is about $50\ ms$. The processing time is the time consumption of the ECUs to operate the cryptographic primitives. The total time consumption includes the processing time of the Secure ECU and the pair of communicating ECUs, the message intervals, and the transmission delay. We perform the key update and distribution process for $10,000$ times on the CAN bus prototype, the average results are shown in Table III. The results shown that the resource of a general ECU is enough for the key update and distribution process. Considering the large amounts of ECUs, the key update and distribution process will inevitably incur overhead. In order to reduce the impact on the performance of the CAN communications, instead of updating and distributing the keys for all the ECUs at one time, the Secure ECU would control this process for each pair of communicating ECUs at regular intervals according to the order of \textit{IDs}.

\subsection{Comparison with Existing Approaches}

\begin{figure}[t]
  \centering
  \includegraphics[width=0.8\linewidth]{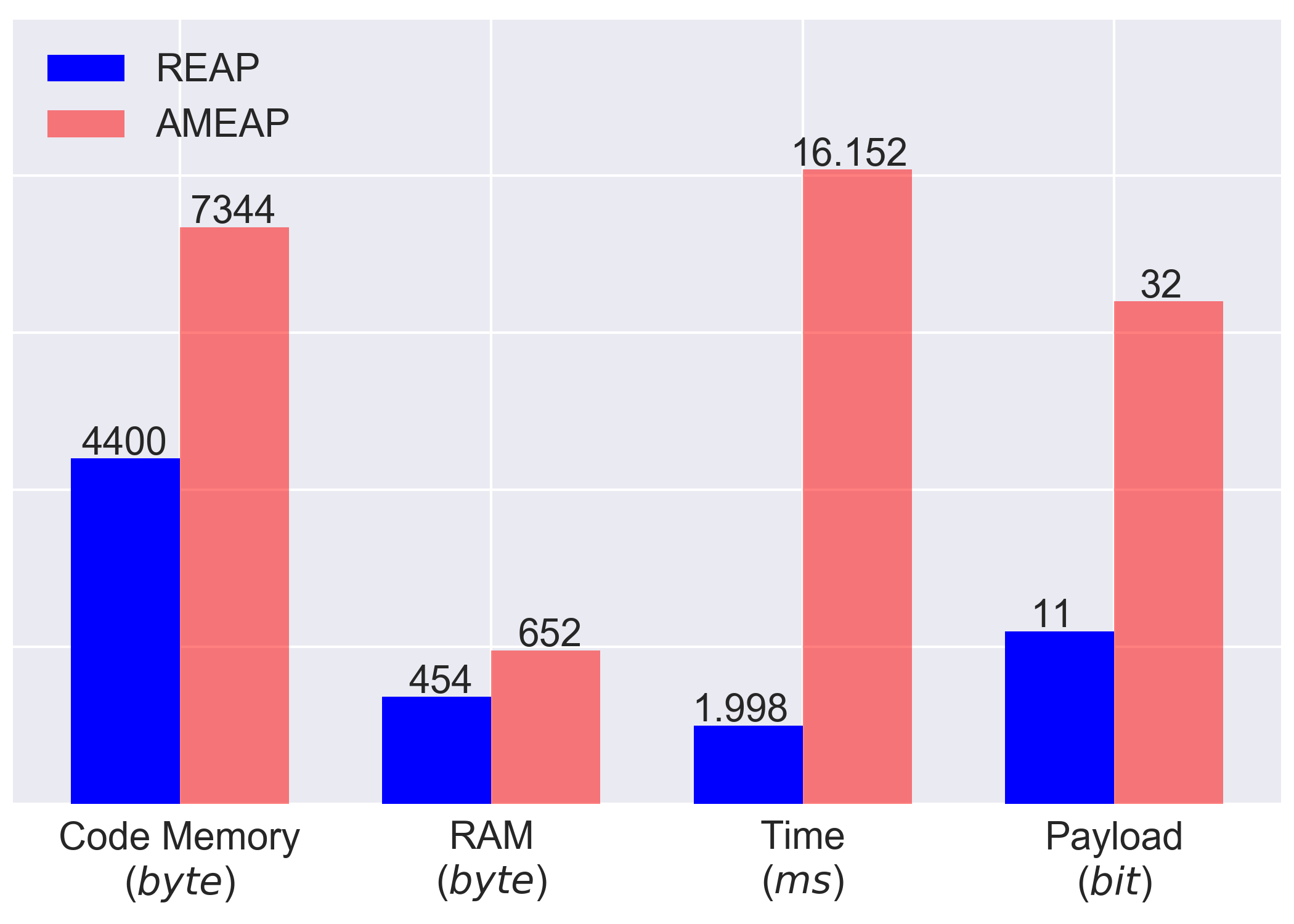}
  \caption{Comparison between LEAP and AMEAP in terms of memory consumption, time consumption for each CAN message, and extra payload for authentication.}
\end{figure}

We take \cite{R13} as a comparison to demonstrate that LEAP is able to provide confidentiality and authentication for the CAN messages with higher efficiency and lower cost. The security protocol of \cite{R13} is denoted as AMEAP because it is based on AES-128 and 32-bit MAC. We control ECU$_1$ to broadcast $10,000$ CAN messages to ECU$_2$ to measure the storage overhead of the two protocols and the average time consumption of the sender and receiver to process a CAN message. As shown in Fig. 9, LEAP consumes less code memory and RAM than AMEAP. It takes $1.998\ ms$ to process one CAN message for LEAP, which is about one eighth of the time consumption of AMEAP. AMEAP needs $32\ bits$ extra payload to insert MAC for the authentication while LEAP only needs $11\ bits$. The results show that LEAP has higher efficiency and lower cost than AMEAP.

\section{Conclusion}

In this paper, we address the issue of protecting the resource-limited in-vehicle networks. The lack of security services in CAN makes it possible for the attackers to track or even control a vehicle remotely, which threats the safety of drivers and passengers. According to the analysis of the existing attack cases, there are four basic attack models based on the attacker's ability and motivation. Taking into consideration the performance and the compatibility of the CAN bus, we propose LEAP to thwart these attacks with less cost and high efficiency. A set of experiments has been conducted on a CAN prototype that contains three ECUs to evaluate the security and performance. Compared with the MAC-based approaches, LEAP can efficiently guarantee the confidentiality and authentication for each CAN message with much less storage overhead and time consumption.

\section*{Acknowledgment}

This work is supported in part of national science foundation of China (Grant No. 61874047 and No. 61376026). Four of us, Zhaojun Lu, Qian Wang, Xi Chen, and Gang Qu, are supported in part by the National Science Foundation under grant CNS1745466 and by AFOSR MURI under award number FA9550-14-1-0351.

\small
\bibliographystyle{IEEEtran}
\bibliography{R1}

\end{document}